\documentclass{article}
\usepackage{graphicx} % Required for inserting images
\usepackage{bm,amsmath,color,dsfont, amssymb}
\usepackage[titletoc,toc,title]{appendix}
\usepackage{graphicx}
\usepackage[section]{placeins}
\usepackage{esint}
\usepackage[hidelinks]{hyperref}
\usepackage[justification=centering]{caption}
\usepackage{cleveref}
\usepackage{subcaption}
\usepackage{float}
\usepackage{multicol}
\usepackage{array} 
\usepackage{tabularx}
\usepackage{tgtermes} %mathptmx %times
\usepackage{relsize}
\usepackage{wrapfig}
\usepackage{siunitx}
\usepackage{bm}
\usepackage{xcolor}
\usepackage[
backend=biber,
style=nature,
]{biblatex}
\usepackage{booktabs}
\usepackage{multirow}

\newcommand{\eps}{\varepsilon}

\usepackage{authblk}
\begin{document}

\title{{The ``double" square-root law: \\ Evidence for the mechanical origin of market impact \\ using Tokyo Stock Exchange data}}
\author[1, 2]{Guillaume Maitrier}
\author[3]{Grégoire Loeper}
\author[4]{Kiyoshi Kanazawa}
\author[5, 2, 6]{Jean-Philippe Bouchaud}

\affil[1]{\textit{LadHyX UMR CNRS 7646, École polytechnique, 91128 Palaiseau, France}}
\affil[2]{\textit{Chair of Econophysics and Complex Systems, École polytechnique, 91128 Palaiseau, France}}
\affil[3]{
\textit{BNP Paribas Global Markets, 20 Boulevard des Italiens, 75009 Paris, France}}
\affil[4]{\textit{Department of Physics, Graduate School of Science, Kyoto University, Kyoto 606-8502, Japan}}
\affil[5]{\textit{Capital Fund Management, 23-25 Rue de l'Université, 75007 Paris, France}
}
\affil[6]{\textit{Académie des Sciences, Paris 75006, France}}

\maketitle

\begin{abstract}
    Understanding the impact of trades on prices is a crucial question for both academic research and industry practice. It is
    well established that impact follows a square-root impact as a function of traded volume. However, the microscopic origin of such a law remains elusive: empirical studies are particularly challenging due to the anonymity of orders in public data. Indeed, there is ongoing debate about whether price impact has a ``mechanical" origin or whether it is primarily driven by information, as suggested by many economic theories. In this paper, we revisit this question using a very detailed dataset provided by the Japanese stock exchange, containing the trader IDs for all orders sent to the exchange between 2012 and 2018. Our central result is that such a law has in a fact microscopic roots, and applies already at the level of single child orders, provided one waits long enough for the market to ``digest'' them. The mesoscopic impact of metaorders arises from a ``double'' square-root effect: square-root in volume of individual impact, followed by an inverse square root decay as a function of time. Since market orders are anonymous, we expect and indeed find that these results apply to any market orders and the impact of synthetic metaorders, reconstructed by scrambling the identity of the issuers, is described by the very same square-root impact law. We conclude that price impact is essentially ``mechanical'', at odds with theories that emphasize the information content of such trades to explain the square-root impact law. 
\end{abstract}
\tableofcontents

\section{Introduction}

The square-root law for price impact is arguably one of the most robust empirical regularities discovered in the last 30 years -- see \cite{loeb1983trading, grinold2000active, almgren2005direct, toth2011anomalous, donier2015million, toth2016square} and \cite{bouchaud2018trades, webster2023handbook} for reviews. It states that when executing a buy (resp. sell) meta-order of total size $Q$, sliced and diced into $N$ child orders of size $q=Q/N$, the price {\it on average} moves up (down) by an amount proportional to $\sqrt{Q}$. Price impact is, quite remarkably, found to be approximately independent of both $N$ and of the total time $T$ needed to achieve full execution \cite{bucci_impact_vol2019}. In other words, provided the participation rate is not too large, average price impact only depends -- to a first approximation -- on the total volume traded $Q$, but not on execution schedule \cite{bouchaud2018trades}. 

Such a square-root dependence, and its apparent universality across a wide variety of markets \cite{bouchaud2018trades}, is surprising and non-intuitive. The classical Kyle model would rather predict a {\it linear} dependence of impact on $Q$, with a slope usually called ``Kyle's $\lambda$'' \cite{kyle1985continuous}. Square-root impact can be interpreted as an effective $Q$-dependent $\lambda$ that {\it diverges} as $Q^{-1/2}$ when $Q \to 0$, as if there was a dearth of liquidity for small volumes \cite{donier2016walras}.  

Several theoretical ideas have been put forth in the literature to explain non-linear impact. Some models predict a concave price impact $Q^\delta$ with $\delta \leq 1$ related to the power-law tail exponent $\alpha$ of the executed volume \cite{gabaix2006} or the power-law tail exponent $\gamma$ of the time autocorrelation of the sign of market orders \cite{farmer2013efficiency,lillo2005theory}. However, as recently shown by Sato and  Kanazawa \cite{kanazawasato2024} using ID-resolved data from the Tokyo Stock Exchange (TSE), the predicted relations between $\delta$ and $\alpha$ or $\gamma$ are not borne out by the data: whereas $\alpha$ and $\gamma$ significantly differ between stocks, exponent $\delta$ remains stubbornly anchored around $\delta=1/2$, i.e. the value corresponding to the square-root law. Another possibility is to invoke the ``propagator model'' which describes the power-law decay of the impact of individual orders \cite{bouchaud2003fluctuations, bouchaud2009markets, bouchaud2018trades}. However, in this case all the concavity comes from a time effect, not from a volume effect -- a possibility clearly ruled out, see \cite{bucci_impact_vol2019}  and Fig. \ref{fig:random_japon} below. 

Yet another line of thought builds upon the classical Glosten-Milgrom model \cite{glosten1985bid} and assumes that metaorders are issued by traders possessing information about future prices, as is traditional in the economics literature \cite{hasbrouck2007empirical, bouchaud2010impact}. Market makers try to detect these informed traders, but do not know their actual participation rate in the total order flow. Using Bayesian arguments, Saddier \& Marsili \cite{saddier2024bayesian} then derive a square-root impact law under certain (weak) conditions and predict that the impact will decay after the end of execution as $t^{-1/2}$ when $t \gg T$.

This last prediction is shared with the Locally Linear Order Book (LLOB) model, which is based on a dynamical theory of ``latent'' liquidity, i.e. volumes that are intended to be exchanged but are not necessarily visible in the order book \cite{donier2015fullyconsistentminimalmodel,bouchaud2018trades}. Plausible assumptions about the dynamics of such liquidity lead to a generic ``V-shape'' in the vicinity of the current price, meaning that local liquidity is vanishingly small -- in line with our above comment on the divergence of Kyle's $\lambda$ for small $Q$'s. Incoming buy (sell) trades then face more and more ``resistance'' as the price moves up (down). The linear increase of liquidity then translates into a square-root behaviour for price impact \cite{donier2015fullyconsistentminimalmodel, donier2016walras}. However, a more precise formulation of the model suggests the existence of two regimes: a small $Q$ regime where the impact is linear and a larger $Q$ regime where the impact is square root. Whereas such a crossover between the two regimes has indeed been reported in Ref. \cite{bucci2019crossover} for very small $Q$'s, the square-root regime sets in for volumes 300 times smaller than predicted by the theory. We will discuss this conundrum more in details below, in the light of our empirical results. 

Let us finally mention that the square-root impact law can also be justified using dimensional arguments coupled with additional plausible assumptions, see \cite{pohl2017amazing, kyle2018market}. 

Despite its critical importance for both financial microstructure and an asset pricing \cite{gabaix2021search, bouchaud2022inelastic}, the very origin of this central phenomenon remains a topic of debate. The universality of the phenomenon suggests a purely mechanical, rather than informational, origin, however this point is 
controversial and at odds with most of the economics literature on the subject, starting with the famous Kyle model \cite{kyle1985continuous}. The aim of the present paper is to give more credence to the ``mechanical hypothesis'' of the square-root market impact law using an ID-resolved data set from the TSE. Our main results are the following:
\begin{itemize} 
    \item The square-root impact law of metaorders is already valid for child orders, provided one waits long enough for the market to digest these orders.
    \item The square-root impact law of metaorders emerges from a ``double'' square-root behaviour: the dependence of the impact of child orders on their individual volume and the inverse square-root relaxation of this impact. 
    \item There does not seem to be anything special about the impact of the child orders of a given metaorder -- in fact all market orders appear to impact prices in the same manner on average. Correspondingly, the impact of synthetic metaorders, reconstructed by randomly scrambling the identity of traders, is identical to the impact of real metaorders. This is our major piece of evidence for a mechanical origin of the square-root law.
\end{itemize}
We also garnered further information shedding light and/or putting constraints on the interpretation of the square-root law:
\begin{itemize}
    \item ``Fast'' traders, for which the holding period is less than a day, represent between 50\% and 60\% of the executed market orders. This includes market makers (HFT) and short term traders. Correspondingly, the fraction of market orders executed against ``fast'' traders represents nearly half of the exchanged volume, a number far too low to vindicate the standard interpretation of the square-root impact within the LLOB framework {\cite{benzaquen2017marketimpactmultitimescaleliquidity}}. Since this empirical fact is inconsistent with the standard LLOB framework, it motivates us to propose a new interpretation of the LLOB framework, as presented in this paper.
    
    \item One can define refill sequences for liquidity providers. One finds that the size of those sequences is also power-law distributed, as was found for metaorders in 
    \cite{sato2023inferring}, following the suggestion of Lillo, Mike \& Farmer \cite{lillo2005theory}.
    \item Once a buy (sell) market order is executed, liquidity providers tend to increase (decrease) their offered price. Such a price degradation however decreases as a power-law of the number of trades already executed, with a prefactor that separates aggressive and wary liquidity providers.   
    \end{itemize}

The outline of the paper is as follows: Section \ref{sec:general_facts} describes the dataset and presents some general facts about execution. Section \ref{sec:micro-meso} investigates the cumulative price impact of child orders and proposes a non-linear propagator to rationalize the empirical results. In Section \ref{sec:synthetic metaorders}, we extend results from the previous Section to all market orders, and we introduce a method for generating synthetic metaorders that are found to follow exactly the same square root law as real metaorders. In Section \ref{sec:liquidity_providers}, we scrutinize the opposite side of market orders by analyzing the behavior of liquidity providers, and we present our conclusions in Section \ref{sec:conclusion}.

\section{Data description and preliminary observations}\label{sec:general_facts}

\subsection{A unique dataset}\label{sub:description_data}

Our study is based on a dataset from the Tokyo Stock Exchange (TSE), provided by the Japan Exchange Group (JPX) for academic purposes only  and already used in \cite{sato2023inferring}. The dataset contains all orders sent to the exchange, with a unique order ID, {a virtual server ID}, the price and type of the order, the volume and price of the best quotes, for all stocks available on the exchange from 2012 to 2018. {Here the virtual server ID is the unit of trading accounts on the TSE. Technically  it is not a membership ID (i.e., the corporate level ID) because any trader may have several virtual servers to avoid the submission-number limit during a fixed interval. However, one can reconstruct an effective trader ID, called the Trading Desk, by properly aggregating those virtual server IDs (see \cite{sato2023inferring, goshima2019trader} for the details). In this paper, the Trading Desks are referred to as trader IDs.}

We focus on the top 100 liquid stocks of the exchange, including 10 ETFs. After anonymizing all assets names (for confidentiality reasons), we only kept orders submitted during continuous double auctions trading sessions: there are two sessions each day in the Japanese market, during 09:00 - 11:30 and 12:30 - 15:00. We discarded orders submitted during the 10 first and last minutes of each sessions, as they might be affected by special conditions. {In the following, we will refer to these two distinct periods as two separate days, with a slight abuse of language.}

We define a {\it metaorder} as a sequence of consecutive market orders (``child orders'') of same sign (buy or sell) submitted by the same trader during a given session. The dataset is unique for two reasons: (i) we have access to a colossal unbiased set of metaorders and (ii) we can analyze traders behaviors as we can assign each orders to a given participant. 

Item (i) is particularly important regarding the claim that impact is an universal mechanism, independent of the type of traders and the information content of the trades \cite{toth2017short, bouchaud2018trades}. Indeed, access to metaorders data is rare (the Ancerno dataset being one exception \cite{zarinelli2015beyond, bucci2018slow, bucci2019crossover}), and most of datasets used in the literature are proprietary and may be plagued by conditioning effects \cite{bouchaud2018trades, webster2023handbook}.\footnote{Note however that CFM data \cite{toth2011anomalous} was acquired in such a way to minimize conditioning effects such as those discussed in \cite{bouchaud2018trades}, ch. 12.3, see also \cite{brokmann2015slow}.} 

Item (ii) represents an interesting opportunity to categorize traders as market makers (MM), high frequency traders (HFT) or low frequency traders, see \ref{sub:slow_fast} and understand better their typical impact on the market. For example, the long term debate about the benefits of HFT for market stability has been dramatically improved with this kind of dataset \cite{hosaka2014analysis, goshima2019trader}. In addition, these identifiers enabled us to analyze the behavior of {\it liquidity providers}, which is only possible if we have access to the ownership of all limit or cancellation orders, see Section \ref{sec:liquidity_providers}. 

\subsection{General stylized facts about metaorders execution}

\begin{figure}[H]
    \centering\includegraphics[width=0.7\linewidth]{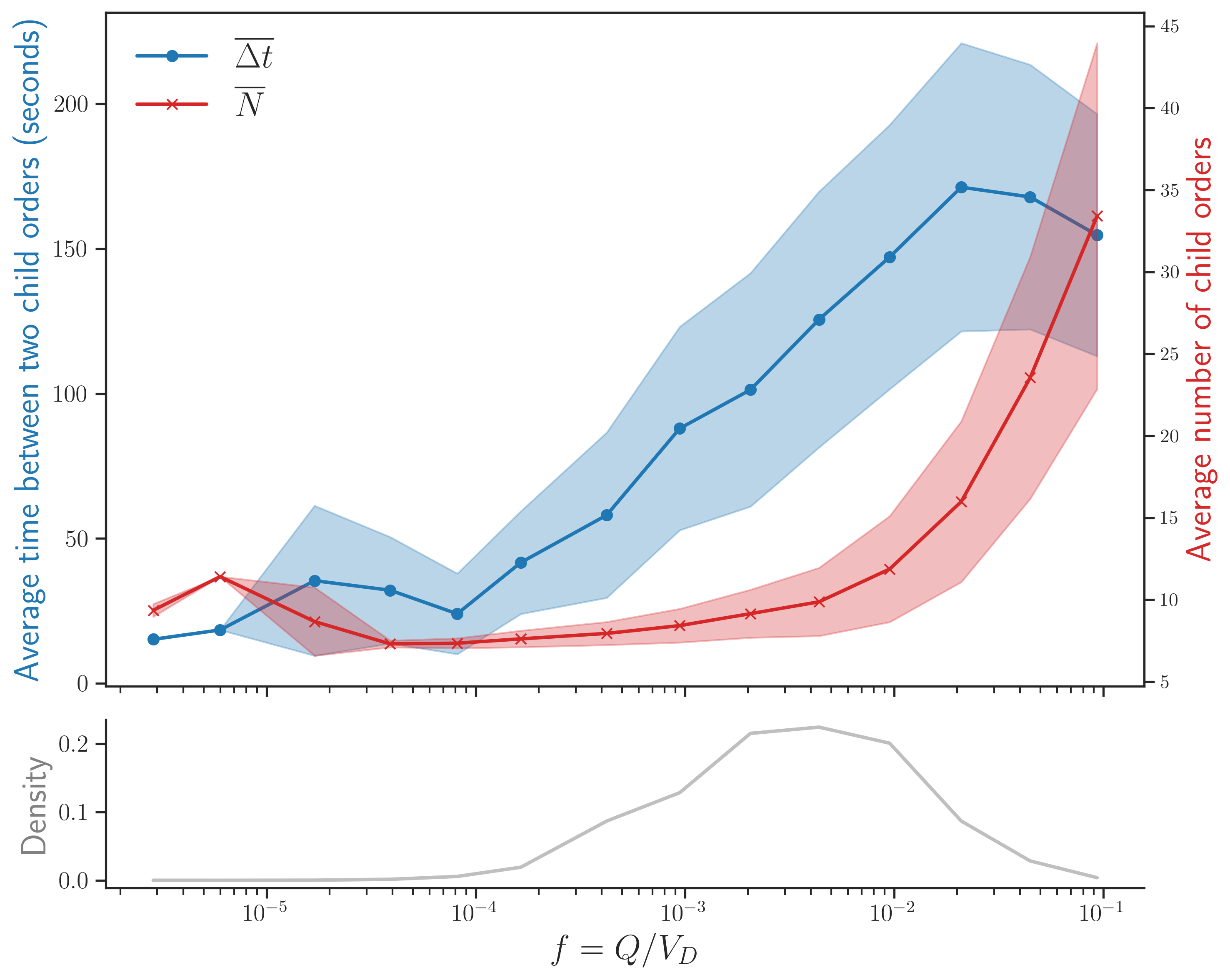}
    \caption{\textbf{Top graph:} The blue points represent the average time  between two child orders of the same metaorder (in seconds), as a function of $f=Q/V_D$. The red points are the average number of child orders as a function of $f=Q/V_D$. The average red curve can be approximately fitted by a power law $f^{0.3}$, for $f\geq 10^{-3}$. The shaded areas represent the corresponding standard deviations. \textbf{Bottom graph:} Distribution of the size of the metaorders, showing a maximum between 0.1\% and 1\% of the daily volume. Results are averaged over the top 24 most liquid TSE stocks.}
\label{fig:distribution_inter_time_vol}
\end{figure}

A metaorder $\omega$ is usually characterized by few metrics: $Q(\omega)$ is the total metaorder size in shares and $N(\omega)$ the total number of child orders, i.e. the number of consecutive market orders of the same sign from the same trader. $T(\omega)$ is the total duration of the execution, $q_i(\omega)$ is the size (in shares) of the $i$th child orders, and $p_i(\omega)$ is the log mid-price just before the time of execution $t_i(\omega)$.

Natural questions that arise (among many others) are:
\begin{itemize}
    \item What is the typical volume $Q$ of metaorders compared to the total daily volume $V_D$?
    \item How does $N$ and $T$ depend on $Q$ on average?
    \item  What is the average execution schedule, i.e. how does the already executed volume $\sum_{t_j \leq t_i} q_j$ depend on $t_i - t_0$?  
\end{itemize} 

We show in Fig. \ref{fig:distribution_inter_time_vol} (bottom graph) the distribution of executed volume fraction $f := Q/V_D$, which shows a broad maximum in the region $f \in [0.1 \%, 1\%]$. Some metaorders correspond to $10 \%$ of the daily volume but they are relatively rare. Similarly, very small metaorders of size $< 0.01 \%$ have a very small probability. The range $f \in [0.1 \%, 10\%]$ for metaorders is typical of firms like, e.g., AQR or CFM \cite{AQR, brokmann2015slow}. 

The average time between child orders ${\Delta t}$ and the average number of child orders are plotted in Fig. \ref{fig:distribution_inter_time_vol} (top graph) as a function of $f=Q/V_D$, where $V_D$  is the executed volume during the day in shares.. We also show as a shaded region the standard deviation of these quantities. One sees that the average time between child orders mildly increases as a function of $f$, ranging from  $25$ secs. for $f=0.01 \%$ to $150$ secs. for $f= 1\%$, before saturating or even slightly decreasing for larger values of $f$. Hence the average execution time $T$ increases slightly faster than $Q$ itself, except perhaps for the largest metaorders.

The increase of $\Delta t$ when $Q$ increases is related to the fact that child orders are more and more aggressive in order to complete execution, so traders wisely wait longer before sending the next one, lest they are detected by market makers. This however becomes difficult for large $f$s, because traders are also attempting to execute their metaorders as quickly as possible. Note that translated into total execution time $T$, these results show that for $f = 0.01 \%$, the typical value of $T$ is $\approx 200$ seconds, whereas for large metaorders with $f = 10 \%$, $T \approx$ 90 minutes. These numbers are however only indicative and the total duration of execution can rise to a full day. 

The typical number $N$ of child orders is around $10$ for $f \lesssim 1\%$ before increasing steeply for larger $f$. This reflects the fact that the available volume at the best quotes is relatively small and if traders want to avoid ``eating into the book'', then the size $q$ of child orders is also limited, which mechanically pushes the number of child orders up when $Q$ increases. Note that typically the volume available at the best quote is around $10^{-4} V_D$ for the most liquid stocks.  

\begin{figure}[H]
    \centering\includegraphics[width=0.6\linewidth]{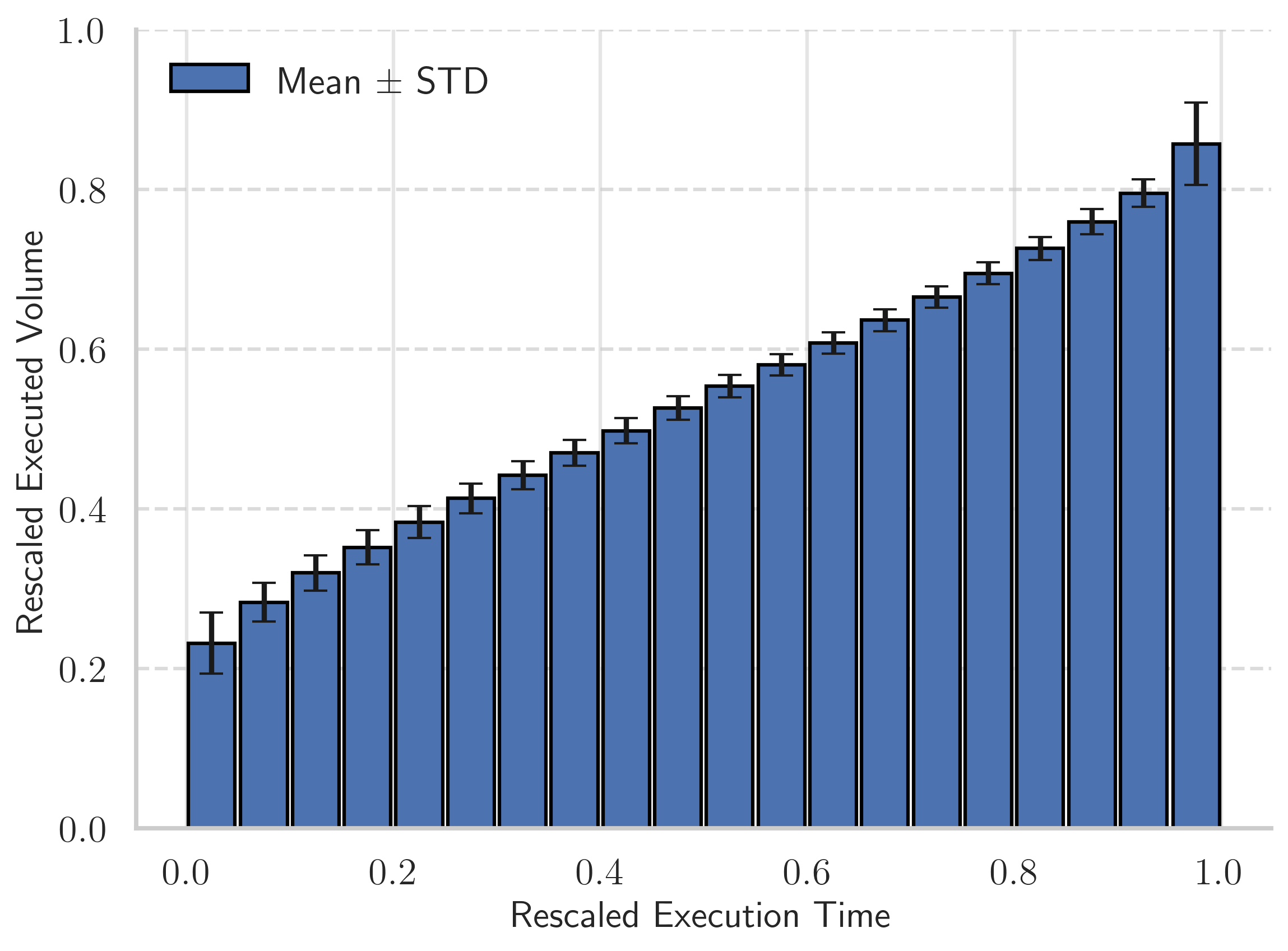}
    \caption{Mean fraction of executed volume of metaorders as a function of the rescaled execution time $(t_i - t_1)/T$. Slight shifts on the $x$-axis come from binning effects. Results from all metaorders in our dataset.}
    \label{fig:enter-label}
\end{figure}

Let us finally turn to the average execution schedule. We plot in Fig. \ref{fig:enter-label} the average executed fraction $\sum_{t_j \leq t_i} q_j/Q$ as a function of the rescaled time since the start of execution $(t - t_1)/T$. One sees a nicely linear average execution profile, suggesting that metaorders are typically executed using a constant trading rate, except possibly at the beginning and at the end of the execution where trades are more aggressive -- although this effect might be dominated by metaorders with a small number of child orders. 

\subsection{Time scales \& Market ecology}\label{sub:slow_fast}

One of the aspects that makes financial markets particularly complex is the wide range of time scales over which traders operate. Indeed, time horizons range from years or decades for institutional investors (pension funds, mutual funds etc) to sub-seconds for market makers. 

These time scales are relevant to understand order flow and liquidity dynamics, and therefore price impact. Indeed, whereas market makers allow orderly trading by acting as intermediaries between final buyers and final sellers, their inventory constraint prevent them from offering ``resistance'' to large buy or sell metaorders. Only slow liquidity can counter-act persistent order imbalance \cite{benzaquen2017marketimpactmultitimescaleliquidity}. 

In this Section, we leverage the identification of traders to classify them into four different classes. We denote by $I_t$ the inventory of a given trader at time $t$, and by $\epsilon_t = \pm 1 $ the sign (buy/sell) of the orders they submit at time $t$. We define the reversal time $\tau$ as the average time between two consecutive orders with different signs from the same trader: 
\begin{itemize}
    \item \textit{Long Term traders}: These traders have long term trading horizon, and typically execute large metaorders. Orders submitted generally have the same sign as their inventories : $I_t \cdot \eps_t > 0$ and the reversal time $\tau$ is typically longer than a trading session.
    \item \textit{Short Term traders}: They can trade high frequency signals that flip sign during a trading session. To exploit their signals, they must however build significant positions $I_t$. Hence we expect them to trade in the same direction for a while, but with a reversal time shorter than a trading session, typically around 30 minutes. In this paper, a ``fast" trader is broadly defined such that their $\tau$ is smaller than the session time as the broadest definition. Otherwise the trader is regarded as a ``slow" trader.
    
    \item \textit{Market Markers}: They provide liquidity to the order book, earning the spread but possibly suffering from price impact. In a way, these participants are the easiest to identify: they are responsible for a large part of trading activity while keeping their inventories $I_t$ close to zero. Their reversal time is typically under the minute.
    \item \textit{Brokers}: They are executing orders on behalf of their clients, so they are trading at high frequency and their inventories can vary greatly across sessions or stocks. It is difficult to assert with certainty that a trader is a broker, as they may resemble the other three categories in a given session.  
\end{itemize}

\begin{figure}[H]
    \centering
    \begin{minipage}{0.5\textwidth}
        \centering
        \includegraphics[width=\linewidth]{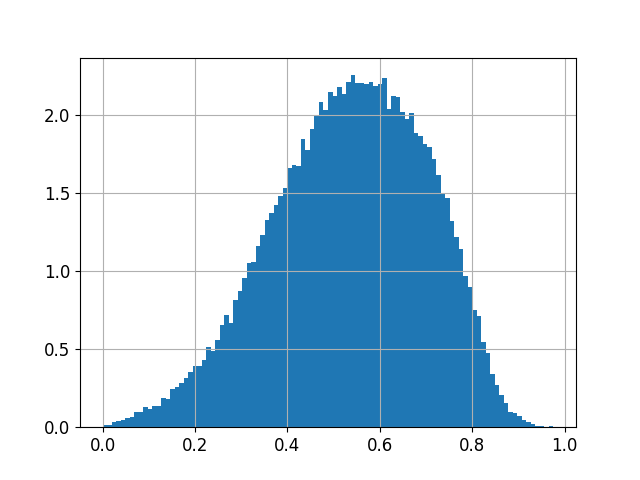}
        \subcaption{${V_{\text{fast}}}/{V_D}$ per session for 100 stocks over 10 years}
        \label{fig:enter-label-1}
    \end{minipage}\hfill
    \begin{minipage}{0.5\textwidth}
        \centering
        \includegraphics[width=\linewidth]{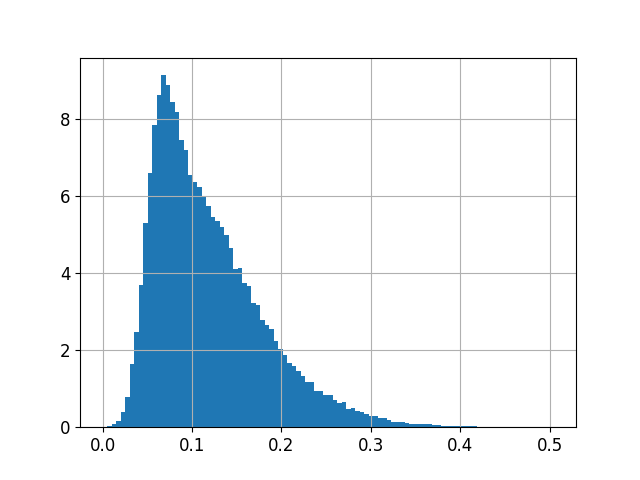}
        \subcaption{${N_{\text{fast}}}/{N_{\text{D}}}$ per session for 100 stocks over 10 years}
        \label{fig:enter-label-2}
    \end{minipage}
    \caption{Histogram of the participation of fast traders to the global trading activity. $V_{\text{fast}}$ is the volume executed by fast trades during a session. $V_{D}$ is the volume of market orders during a session. $N_{\text{fast}}$ is the number of fast traders and $N_{D}$ is the total number of traders participating to a given session. We used each session for each stock in our database (around 40,000 sessions).}
    \label{fig:slow_fast_hist}
\end{figure}

Using the reversal time $\tau$ as a criterion to separate ``fast'' and ``slow'' traders, we compute the contribution of the two categories to the trading activity using either the fraction of the market order volume executed by fast traders, $V_{\text{fast}}/V_D$ or the relative fraction of the number of fast traders, $N_{\text{fast}}/N_D$. {$N_{\text{fast}}$ is the number of ``fast'' traders in a given session, whereas $N_D$ is the total number of traders having traded at least once during the session.} We show in Fig. \ref{fig:slow_fast_hist} the histogram of these ratios, computed over all sessions and all stocks of our database. Whereas the most probable value of $N_{\text{fast}}/N_D$ is around $8 \%$, the most probable value of the ratio $V_{\text{fast}}/V_D$ is between $50 \%$ and $60 \%$. Therefore, the contribution of ``slow'' market orders to total volume is roughly one-half. This finding is in line with \cite{Ohyama}. One can also determine the fraction of market orders executed against fast traders, which is found to be in the range  $60 \%$ to $70 \%$.

The conclusion of this study is that while fast traders are dominant in terms of volume, the contribution of slow volumes to trading activity actually of the same order of magnitude.  This finding is important since the standard interpretation of the square-root impact law in terms of latent liquidity \cite{donier2015fullyconsistentminimalmodel, benzaquen2017marketimpactmultitimescaleliquidity} assumes that slow volumes are a factor $\sim 300$ times smaller than $V_D$ \cite{bucci2019crossover}, which is certainly not the case here. In the next Section, we will revisit the empirical evidence for square-root impact and propose a new interpretation of the LLOB model.

\section{Square-root impact: micro-scales \& meso-scales}\label{sec:micro-meso}

As recalled in the introduction, there is overwhelming empirical evidence for the square-root impact law for metaorders, which has again been confirmed in great detail in \cite{kanazawasato2024} for the TSE, using the very same data set as here. More precisely, the square-root impact law states that
\begin{equation}\label{eq:sqrt_law}
\boxed{
    \mathcal{I}(Q) : = \mathbb{E}[\Delta p\cdot \epsilon \mid Q] = Y \sigma_D\sqrt{\frac{Q}{V_D}}
    }
\end{equation}
where $\epsilon$ is the sign of the metaorder of total size $Q$, $\Delta p$ is the log mid-price change between just before the first and just after last child order and $\sigma_D$ and $V_D$ are the contemporaneous daily volatility and exchanged volume. Note again that $\mathcal{I}(Q)$ is independent of the execution time $T$ (see for example \cite{bucci_impact_vol2019} and Fig. \ref{fig:random_japon} below). {In the rest of the paper, we will use $({p_{\rm{high}} - p_{\rm{low}}})/{p_{\rm{open}}}$ as a proxy for $\sigma_D$}. 

In this Section we attempt to dissect the square-root law into more microscopic components, which sheds further light into its origin and leads to a new interpretation of the Latent Liquidity model. 

\subsection{The ``double'' square-root impact of child orders}

\begin{figure}[H]
    \centering
    \includegraphics[width=0.8\linewidth]{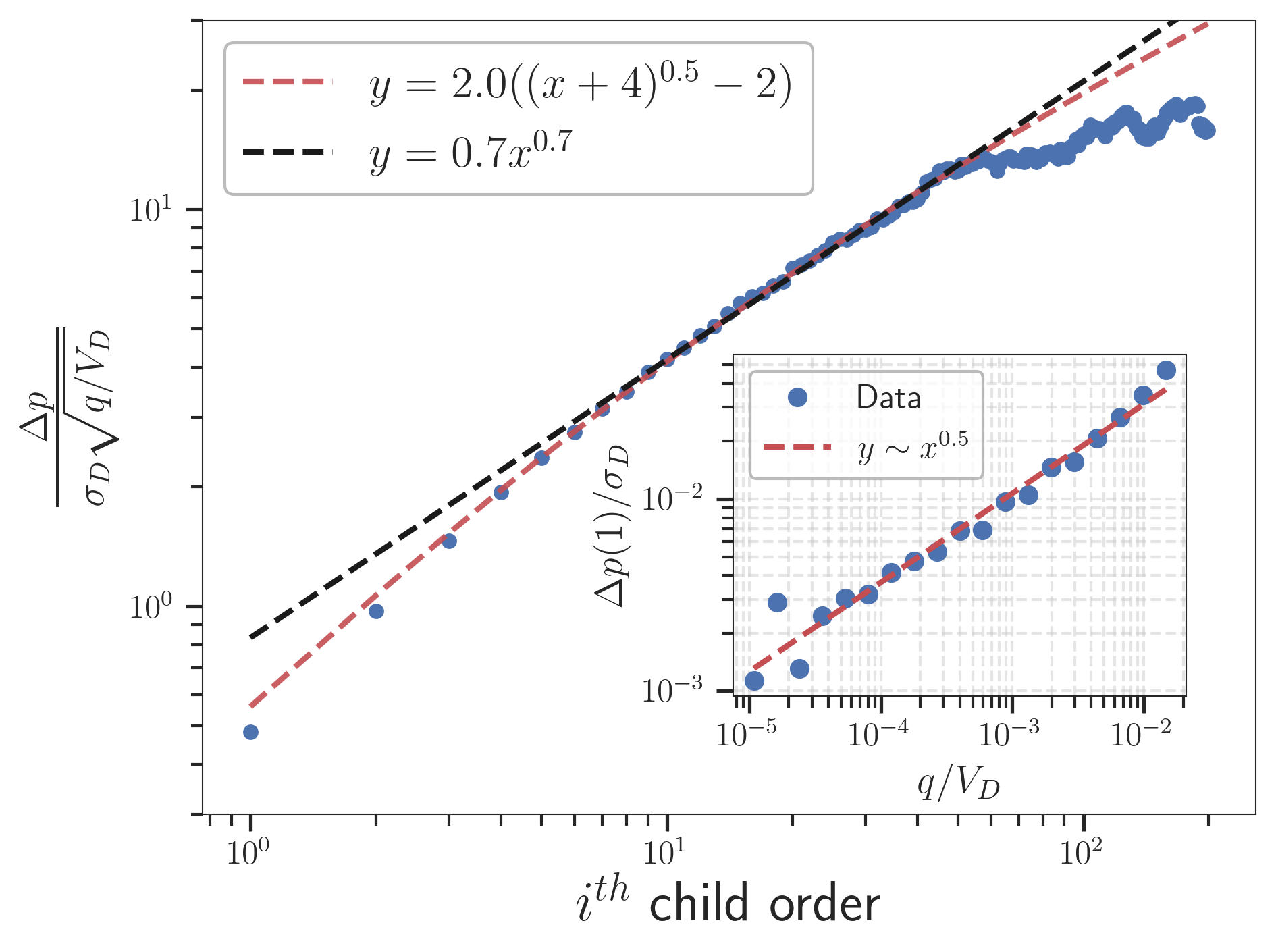}
    \caption{Average price profile during the execution of a metaorder. The {vertical axis} represents the cumulative impact of child orders, rescaled by the daily volatility and the square root of the relative volume of the child order $q$. These profiles are obtained averaging over the top 8 most liquid stocks of our dataset. We show with red dotted line the function $\left(\sqrt{i+i_0} - \sqrt{i_0}\right)$ and in black a pure power-law fit $i^{0.7}$. Inset: Average impact of the first child order $\mathbb{E}[\Delta p(1) \cdot \epsilon]$ as a function of its size $q$ rescaled by daily volume $V_D$, demonstrating the fact that the square-root law is in fact valid at the level of child orders. }
    \label{fig:linear_LLOB}
\end{figure}

Instead of measuring the impact $\mathcal{I}$ of full metaorders, one can measure the partial impact $\mathcal{J}(q,i)$, i.e. the average price difference $\Delta p_i$ between just before child order $i+1$ and just before child order $1$, {conditional on} the size $q$ of child orders and the rank $i$. We find that to a good approximation that impact is proportional to {\it both} $\sqrt{q}$ and $\sqrt{i}$: 
\begin{equation} \label{eq:sqrt_child}
\boxed{\mathcal{J}(q,i) := \mathbb{E}[\Delta p_i \cdot \epsilon {\vert  q,i}] \propto \sigma_D\sqrt{\frac{q}{V_D}} \, \Big(\sqrt{i+i_0} - \sqrt{i_0}\Big)}
\end{equation}
with $i_0 \approx 4$, see Fig. \ref{fig:linear_LLOB}. We show in the inset that the $\sqrt{q}$ dependence on the volume of child orders holds very well for $i=1$, but we have checked such a dependence for other values of $i$ as well. This demonstrates that the square-root law {\it already holds at the level of child orders}, provided one waits long enough after the execution, see below. Eq. \eqref{eq:sqrt_child} is the central result of this paper. 

Several remarks are in order: 
\begin{itemize}
    \item The fit is very good up to $i=50$, beyond which another regime appears to set in, where impact saturates. This however only concerns a small fraction of large metaorders, for which conditioning effects should be taken into account (e.g. large prevailing liquidity at the opposite best, see \cite{patzelt2018universal}). An alternative interpretation is that such large metaorders are detected by the market, triggering the influx of opposing limit orders. Impact saturation for large $Q$ has been reported elsewhere as well, see e.g. \cite{zarinelli2015beyond, bucci2019crossover}. 
    \item When fitting with a more general power-law $(i + i_0)^{1-\beta} - (i_0)^{1-\beta}$, the optimal value of $\beta$ is found to be $0.48$, i.e. very close to a square-root.\footnote{The exponent $\beta$ is defined as the decay exponent of the propagator, as $|t_i - t_j|^{-\beta}$, see \cite{bouchaud2003fluctuations, bouchaud2018trades}.} Alternatively, imposing $i_0=0$ yields $1-\beta = 0.7$, but the fit is clearly worse for small times, see Fig. \ref{fig:linear_LLOB}.
    \item When $i \gg i_0$, one finds that the temporal profile of the impacted price behaves as $\sqrt{i}$, a result already reported in \cite{moro2009market, zarinelli2015beyond, donier2015million}. 
    \item When $i=N$ and using $Q=qN$ one finds, {after injecting $i=N$ in Eq. \eqref{eq:sqrt_child} and factorizing,} 
    \begin{equation} \label{eq:sqrt_Q}
        \mathcal{I}(Q) \propto \sigma_D\sqrt{\frac{Q}{V_D}} \left(\sqrt{1+i_0/N} - \sqrt{i_0/N}\right),
    \end{equation}
    allowing one to recover exactly the square-root impact law, Eq. \eqref{eq:sqrt_law}, up to a weakly varying $N$-dependent factor that increases from $0.31$ to $1$ as $N$ goes from $2$ to $\infty$ (when $i_0=4$).  
\end{itemize}

\subsection{A non-linear propagator model}

The above results can be summarized within the framework of a non-linear propagator model \cite{bouchaud2003fluctuations, bouchaud2018trades}, where the impact of child order $j$ measured at time $t_i > t_j$ is proportional to $\sqrt{q_j/(t_i-t_j + s_0)}$, as predicted both by the LLOB model \cite{donier2015fullyconsistentminimalmodel} and by the Bayesian theory of Ref. \cite{saddier2024bayesian}. Indeed, from such an expression one gets:
\begin{equation}\label{eq:sqrt_child_partia}
    \mathcal{J}(q,i) \propto \sqrt{q} \sum_{t_j \leq t_i} \frac{\sqrt{\Delta t}}{\sqrt{t_i - t_j + s_0}} \approx 2\sqrt{q} \left(\sqrt{i+i_0} - \sqrt{i_0}\right), \qquad s_0 \equiv i_0 \Delta t,
\end{equation}
where we have assumed for simplicity that $q_j = q$, $\forall j$ and $t_j \approx j \Delta t + t_0$, with $\Delta t$ the time between two consecutive child orders. We have furthermore approximated the discrete sum over $j$ by an integral over $t_j$. 

Such an interpretation however appears to violate the ``diffusivity'' condition derived in \cite{bouchaud2003fluctuations, jusselin2020no}, which relates the decay of the propagator with the decay of the autocorrelation of the sign of the trades. Superficially, a propagator decay in $(t_i-t_j)^{-1/2}$ should lead to strongly mean reverting prices, at odds with the diffusive nature of prices. A way out of this conundrum will be presented in a forthcoming paper \cite{ustocome}. Note that the above non-linear propagator model precisely saturates the no-arbitrage bound derived by Gatheral in \cite{gatheral2010no}. 

The most striking result of the previous Section is that the square-root law appears to hold already at the level of child orders. This is not in line with the standard ``mesoscopic'' interpretation of the Latent Liquidity model \cite{donier2015fullyconsistentminimalmodel}, which, as we alluded to before, would require the fraction of slow volume to be much smaller than what we reported in Section \ref{sub:slow_fast}. 

We are thus led to the conclusion that the latent liquidity idea must in fact operate already at the micro level. Whereas the revealed order book contains primarily limit orders posted by market makers, the final sellers' or buyers' price would be distributed according to a locally linear profile, as predicted by the LLOB theory \cite{donier2015fullyconsistentminimalmodel} -- which, as a reminder, only relies on minimal assumptions, in particular on the diffusive nature of prices. 

So the scenario would be as follows: once the incoming buy (sell) has been executed against a market-maker, a ``hot potato'' game starts between market-makers until the order is finally digested by a final seller (buyer). In order to find such a final seller (buyer) the price must on average move by an amount $\delta p$ such that $\Gamma (\delta p)^2/2 = q$, where $\Gamma$ is the slope of the latent liquidity. Once this is achieved, the price tends to revert back as $1/\sqrt{t - t_i}$, as predicted by the LLOB model, see \cite{donier2015fullyconsistentminimalmodel, bouchaud2018trades}, but in disagreement with the predictions of the propagator model \cite{bouchaud2003fluctuations}. This discrepancy will be discussed further in \cite{ustocome}.

If the above interpretation is correct, however, it should hold for arbitrary market orders. Since all orders are equivalent, they should lead to the same average impact on prices (as was indeed found in \cite{toth2017short}). In other words, one should see a $\sqrt{q}$ impact for single market orders provided one waits long enough for the ``hot potato'' game to be completed. This is what we test in the next Section, which then opens up the question of reconstructing synthetic metaorders from a list of consecutive market orders that do not necessarily belong to the same ID. The anonymity of market orders suggests that, in certain conditions, the very same square-root impact law $\mathcal{I}$ given by Eq. \eqref{eq:sqrt_law} should also hold for synthetic metaorders. This indeed turns out to be the case, as we discuss now. 
\newpage
\section{From single market orders to synthetic metaorders}\label{sec:synthetic metaorders}

\subsection{The impact of single public market orders}

We want to understand the behavior of the price after a buy (sell) market order. Clearly, if the volume of the market order is less than the prevailing volume at the opposite best, the immediate impact is zero. However, as time goes by, one very quickly sees that impact grows and becomes approximately given by $\sqrt{q}$, whether or not immediate impact is zero (the full temporal aspects will be explored in more details in \cite{ustocome}). 

\begin{figure}[H]
    \centering
    \begin{minipage}{0.5\textwidth}
        \centering
        \includegraphics[width=\linewidth]{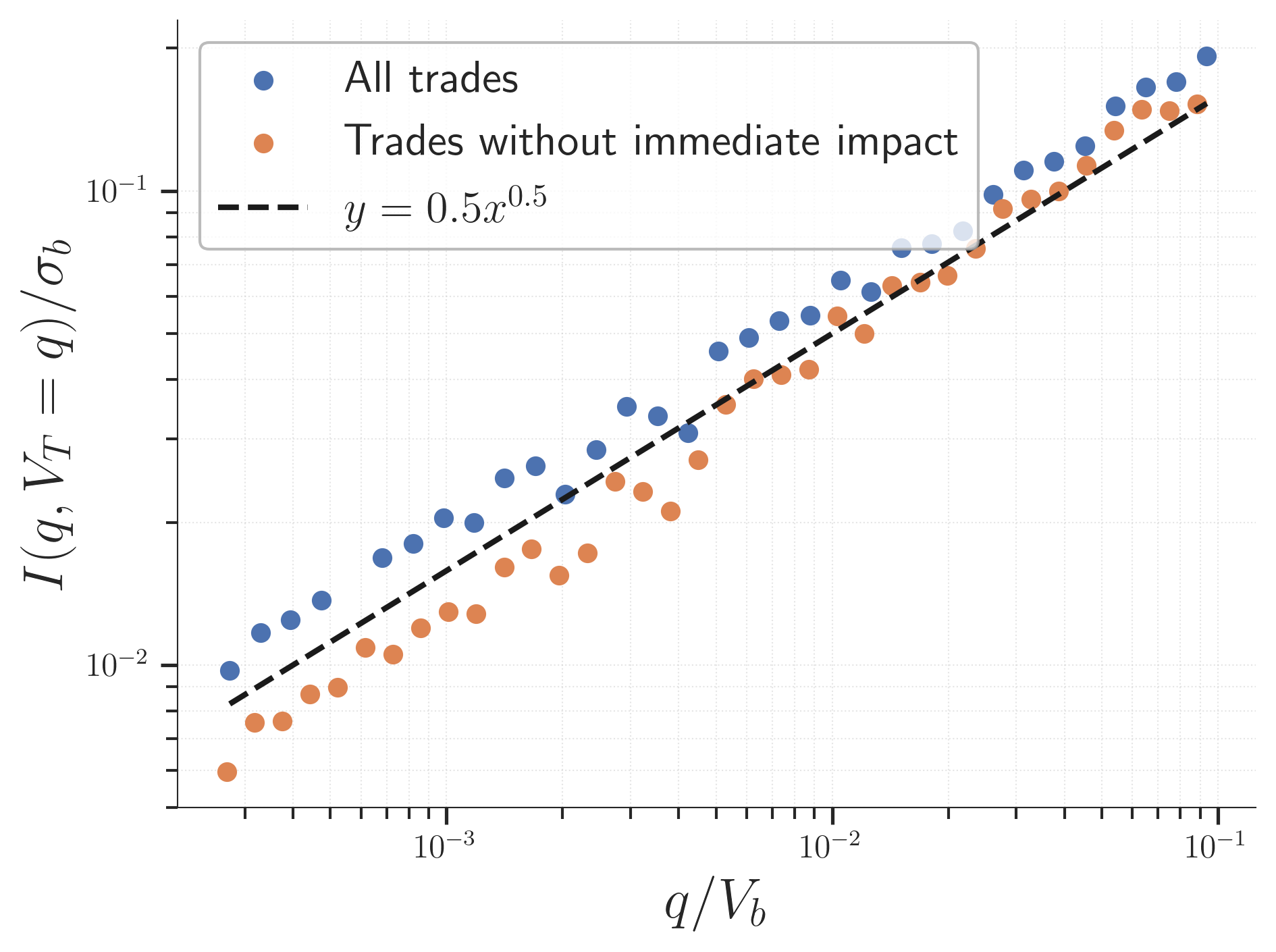}
        \label{fig:enter-label-3}
    \end{minipage}\hfill
    \begin{minipage}{0.5\textwidth}
        \centering
        \includegraphics[width=\linewidth]{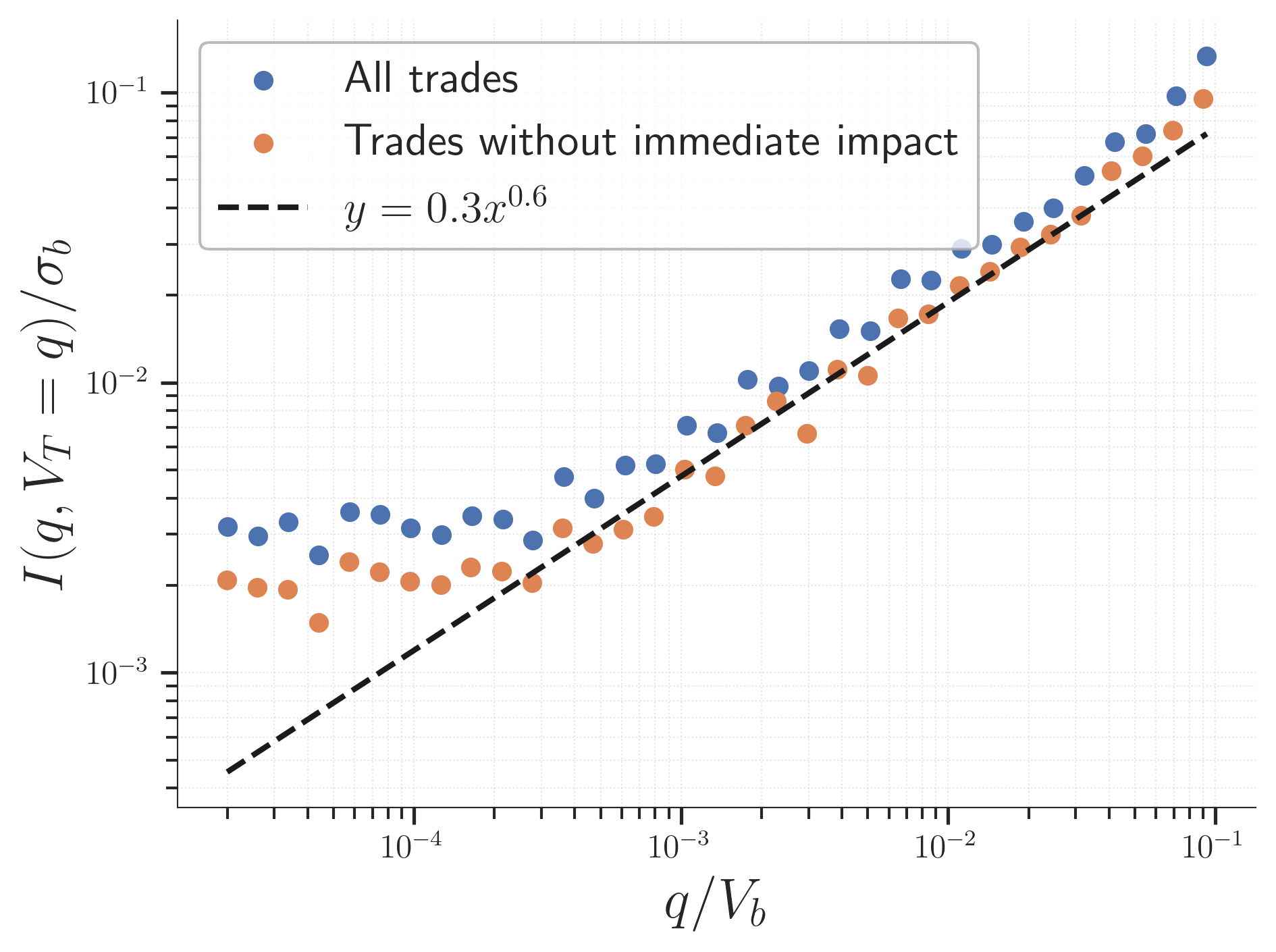}
        \label{fig:enter-label-4}
    \end{minipage}
     \caption{Impact of public market orders measured after a volume time equal to the size $q$ of the market order itself. {$\sigma_b$ and $V_b$ are the average volatility of volume of the 15 minute bin to which the order belongs.} The blue points represents the average impact of all trades, while the orange points represents trades that have no immediate impact, i.e. such that $q$ is smaller than the prevailing volume at the opposite best. In this way, we can see that the impact indeed builds up over time. Left and right graphs correspond to two typical stocks, one showing a nearly perfect $\sqrt{q}$ behaviour for all $q$ (see dashed black line). Note that the right graph exhibits a plateau for very small $q$'s.} 
    \label{fig:single_MO}
\end{figure}

More precisely, we show in Fig. \ref{fig:single_MO} the impact of a single market order as a function of $q$ for two typical stocks of the TSE, after waiting for a volume time equal to $q$ itself -- i.e. after the market has traded the same quantity as the initial market order. We see that independently of whether or not the initial market order has an immediate impact, the overall behaviour is compatible with an impact growing as $\sqrt{q}$. {To remove intraday seasonality effects, we first determine the average intraday profile of volatility $\sigma_b$ and executed volume $V_b$, using yearly data, computed on each 15 minutes bins. Then, we rescale impact and volume of market orders by, respectively, $\sigma_b$ and $V_b$ corresponding to the 15-minute bins of the day to which these orders belong.} 

For some stocks, a plateau regime however appears for very small $q$, perhaps related to tick size effects (see Fig. \ref{fig:single_MO}; right graph). 

\subsection{Synthetic metaorders}

Since arbitrary market orders seem to all behave similarly, independently of the metaorder they belong to, we made the following numerical experiment that confirms the non-linear propagator interpretation of Eq. \eqref{eq:sqrt_child} for metaorders. We construct a new dataset of synthetic metaorders, by randomly shuffling traders ID and distributing them to market orders while we keep the historical market order flow, i.e. by randomly reordering real traders' IDs \footnote{{The shuffling was based on the Fisher-Yates algorithm. In other words, we collect all the market orders within the same session for a specific stock and simply shuffle only trader IDs. Specifically, we apply the function numpy.random.shuffle in Python to the DataFrame column containing the trader IDs}}. This preserves the initial frequency distribution of traders: i.e. some of them appear many times whereas others are trading less frequently.  

\begin{figure}[H]
        \centering
        \begin{minipage}{0.5\textwidth}
            \centering
            \includegraphics[width=\linewidth]{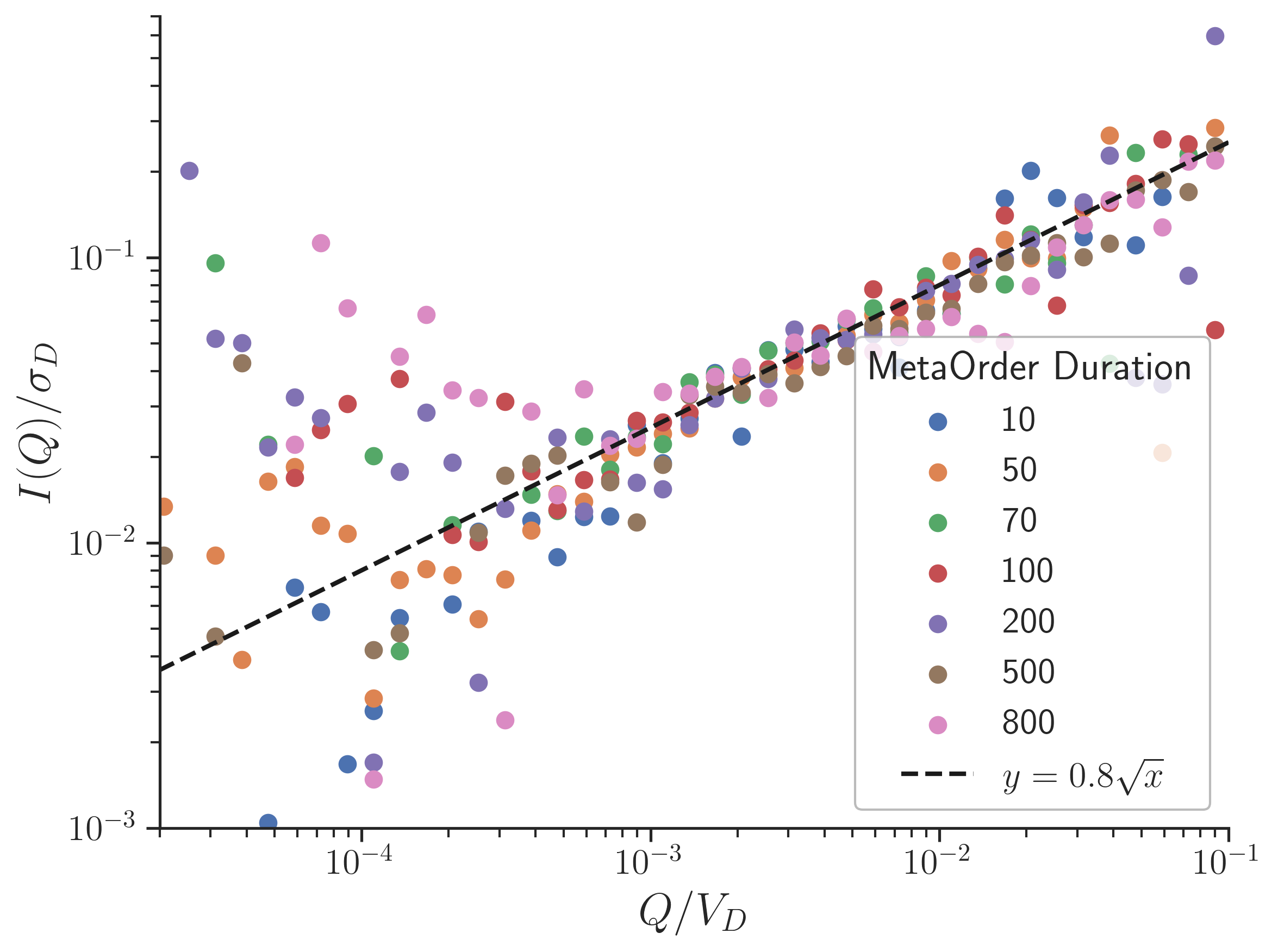}
        \end{minipage}\hfill
        \begin{minipage}{0.5\textwidth}
        \centering
        \includegraphics[width=\linewidth]{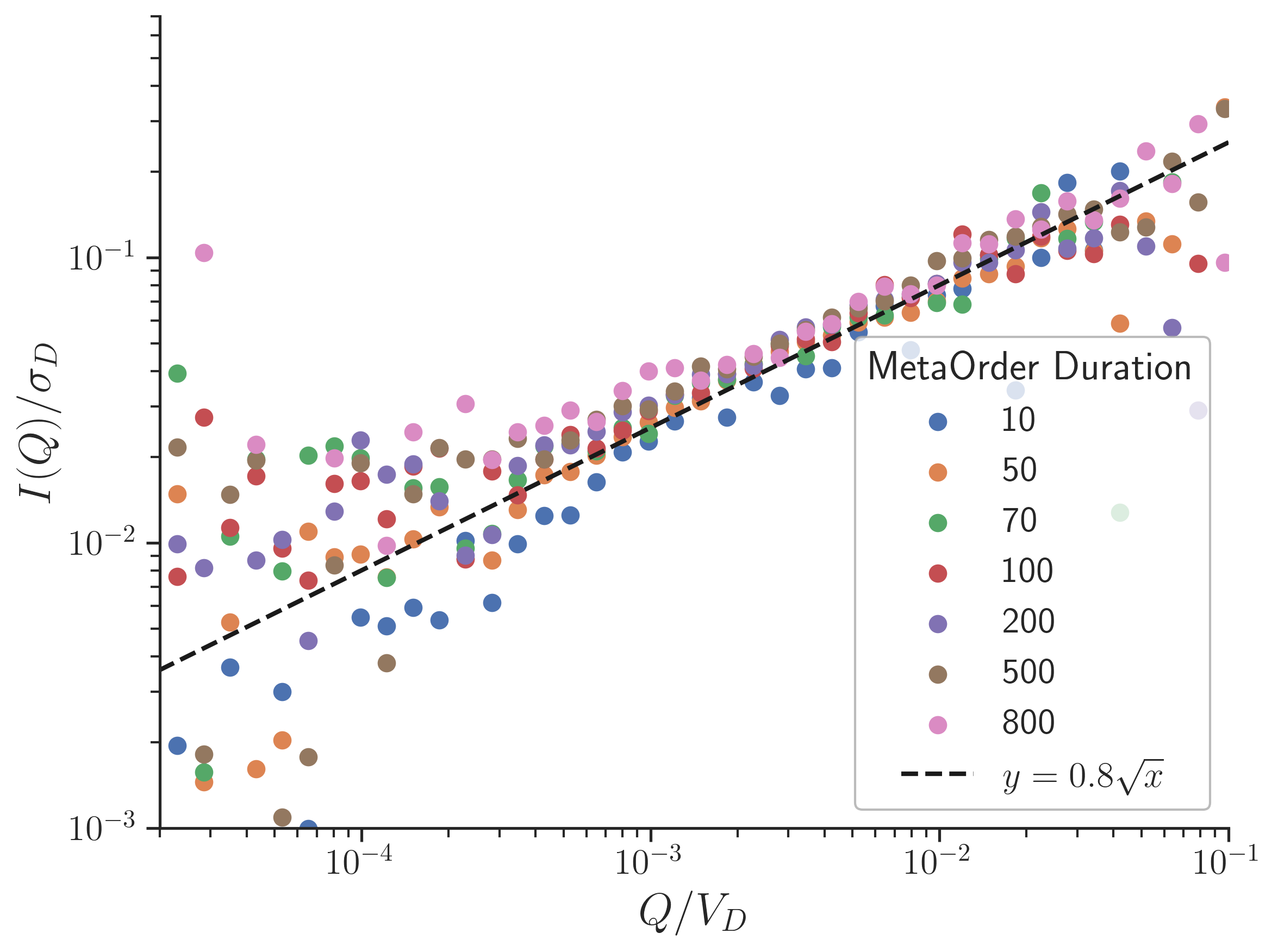}     \end{minipage}
    \caption{\textbf{Left:} Impact $\mathcal{I}(Q)$ of real metaorders as a function of their rescaled size $Q/V_D$ for a typical Japanese Stock, using data from 2012 to 2018. The color of the dots correspond to different total execution time $T$, expressed in seconds. 
    \textbf{Right:} Impact $\mathcal{I}(Q)$ of synthetic metaorders for the same stock, obtained via ID reshuffling. ID shuffling consists in a random permutation of historical trader IDs, preserving the frequency of apparition. Note that the {vertical and horizontal scales} are the same in the two plots: the square-root fit is exactly the same as for real metaorders. From the legend, one can also clearly see that $\mathcal{I}(Q)$ is independent of $T$ \cite{bucci_impact_vol2019}.} However, it is worth noting that synthetic metaorders are generally smaller in size compared to real ones.
      \label{fig:random_japon}
\end{figure}

We then use the same  method as in \ref{sub:description_data} to define metaorders as a consecutive sequence of trades of the same sign associated to the same new trader ID (that has been reshuffled). We obtain synthetic metaorders, that start and end at different times as the original (true) metaorders. Hence, any information associated with these metaorders is at least partially lost. Still, as shown in Fig. \ref{fig:random_japon}, we recover exactly the same square-root impact function as for the original metaorders! Note in passing that these graphs show once again that the square-root impact law only depends on the volume $Q$ of the metaorder (either real or synthetic) but not of the execution time $T$, see \cite{bucci_impact_vol2019, bouchaud2018trades, kyle2018market}. We have tested different constructions of synthetic metaorders {on different stocks, from the Paris and London Stock Exchange}, with similar results -- a more detailed discussion will be presented in \cite{ustocome}.
{Note that the preservation of the impact law under reshuffling provides further evidence that short-term impact should be decoupled from alpha (i.e., predictive signals). This is consistent with the fact that the typical time horizon over which traders seek alpha is significantly longer than the execution timescale of metaorders. This is particularly true in this study, as metaorders are defined within trading periods. Similar observations have been made, for instance with CFM’s trades—where the square-root law remains valid—or in the ANcerno dataset, see \cite{toth2011anomalous, toth2016square, bucci2018slow}}

\subsection{Discussion}

The results of the previous two sub-sections strongly suggest a purely ``mechanical'' interpretation of the square-root impact law, based on a time decaying $\sqrt{q/t}$ impact of single market orders, independently of their association with a specific metaorder (since trader ID's can be scrambled without affecting the results). These findings are difficult to reconcile with theories explaining the square-root law based on information, such as in \cite{gabaix2006}, or on the detection by the market of the beginning of new metaorders, such as in \cite{farmer2013efficiency, saddier2024bayesian}.

Those results are, on the other hand, perfectly in line with the fact that, due to anonymity, all market orders -- even uninformed ones -- should play an equivalent role and should on average impact prices similarly \cite{bouchaud2010impact}. This was already noted in \cite{toth2017short} by comparing the impact of CFM market orders with non-CFM market orders, and even more convincingly in an unpublished specifically designed 2010 experimental campaign with totally random market orders.  

\section{The other side of market orders: liquidity providers}
\label{sec:liquidity_providers}
\subsection{Refill sequences}

Whereas the long-term correlation of market orders is well documented \cite{bouchaud2003fluctuations, lillo2005theory, bouchaud2009markets} and mainly attributed to the order splitting of large metaorders \cite{lillo2005theory, sato2023inferring}, our dataset also allows us to study the splitting strategy of liquidity providers. 

\begin{figure}[H]
    \centering
    \includegraphics[width=\linewidth]{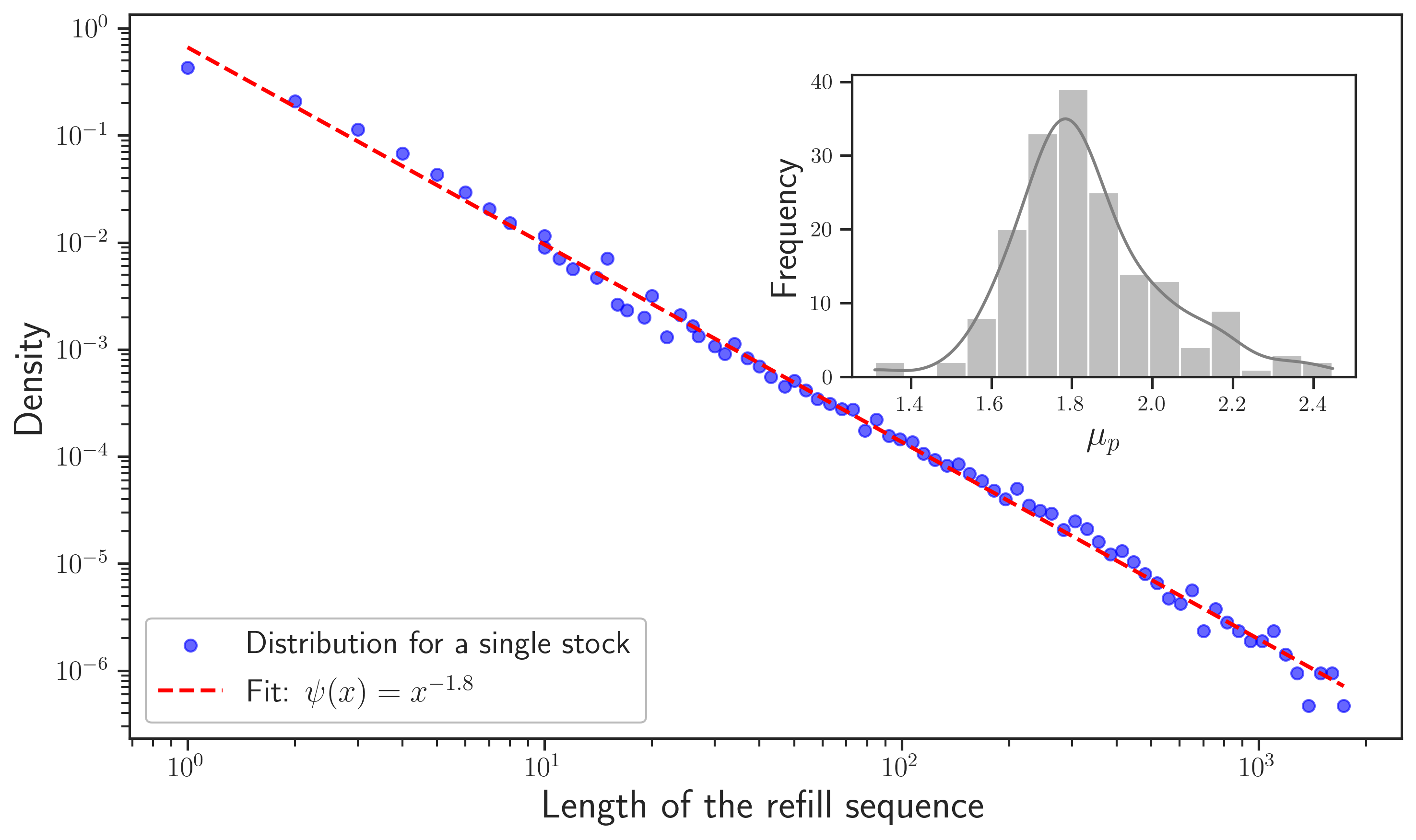}
    \caption{Distribution of the length $n$ of refill sequences for a given stock of the TSE. A power-law distribution fits the data very well: $\psi(n) \sim n^{-\mu_p}$. \textbf{Inset:} Distribution of $\mu_p $ across the different stocks of our dataset, regressed independently.}
    \label{fig:LMF_Provider}
\end{figure}

To do so, we simplify the problem by restricting to the set of \textit{filled} limit orders, i.e. limit orders that have been placed in the order book and subsequently executed by another participant. Thus, as for liquidity takers, one can aggregate those filled limit orders into ``refill sequences'', i.e. sequences of consecutive filled limit orders of same sign submitted by the same trader during a trading session. Given the splitting behavior of liquidity takers, market makers/liquidity providers face a sign-correlated succession of market orders. It is thus likely that the executed limit orders flow will be also be persistent, and lead to a power-law tail in the size distribution of the  refill sequences, as we indeed confirm empirically, see Fig. \ref{fig:LMF_Provider}. We show there a power-law fit of the distribution of the number $n$ of child orders associated with refill sequences, as $\psi(n) \propto n^{-\mu_p}$ with $\mu_p$ in the range $[1.4, 2.4]$ depending on the considered stock. This power-law decay echoes the Lillo-Mike-Farmer distribution of child market orders, although it is not expected to mirror it exactly since different traders provide liquidity to the same incoming metaorder -- see \cite{eisler2012price} for a related discussion.  

It should however be emphasized at this point that the boundary between liquidity provider  and liquidity taker is somewhat blurred. Except for specific cases, the large majority of market participants use a mix of limit and market orders to acquire or sell shares. For example, it is quite common to see participants adding liquidity at the bid (ask) while sending market orders at the ask (bid). Large funds like AQR declare that most of their executed volume is though limit orders \cite{AQR}.

\subsection{Strategic behaviour of liquidity providers}

As liquidity providers get executed on the ask (bid) side, they tend to increase (decrease) their next limit order such as to (i) control their inventory as the next trade will be biased towards the bid (ask), (ii) protect themselves against being picked up by making the next trade less favorable for the buyer (seller). This is often called ``skewing'' in the market making jargon, and/or {(iii) ask for a better price in the case demand for liquidity is persistent.}

Hence we expect the next limit order to be executed at a higher (lower) price, i.e. 
\[
\mathcal{K}(i):=\mathbb{E}\left[\epsilon \cdot \frac{p_{i+1} - p_i}{\sigma_D} \mid i \right] > 0,
\]
where $\epsilon$ is the sign of the executed market order, and $p_i$ is the log-price at which the $i$th child of a refill sequence is filled. We find that the ``refill function''  $\mathcal{K}(i)$ depends only weakly of the volume of the filled limit order, probably due to strategic liquidity provision.

\begin{figure}[H]
    \centering
    \includegraphics[width=0.7\linewidth]{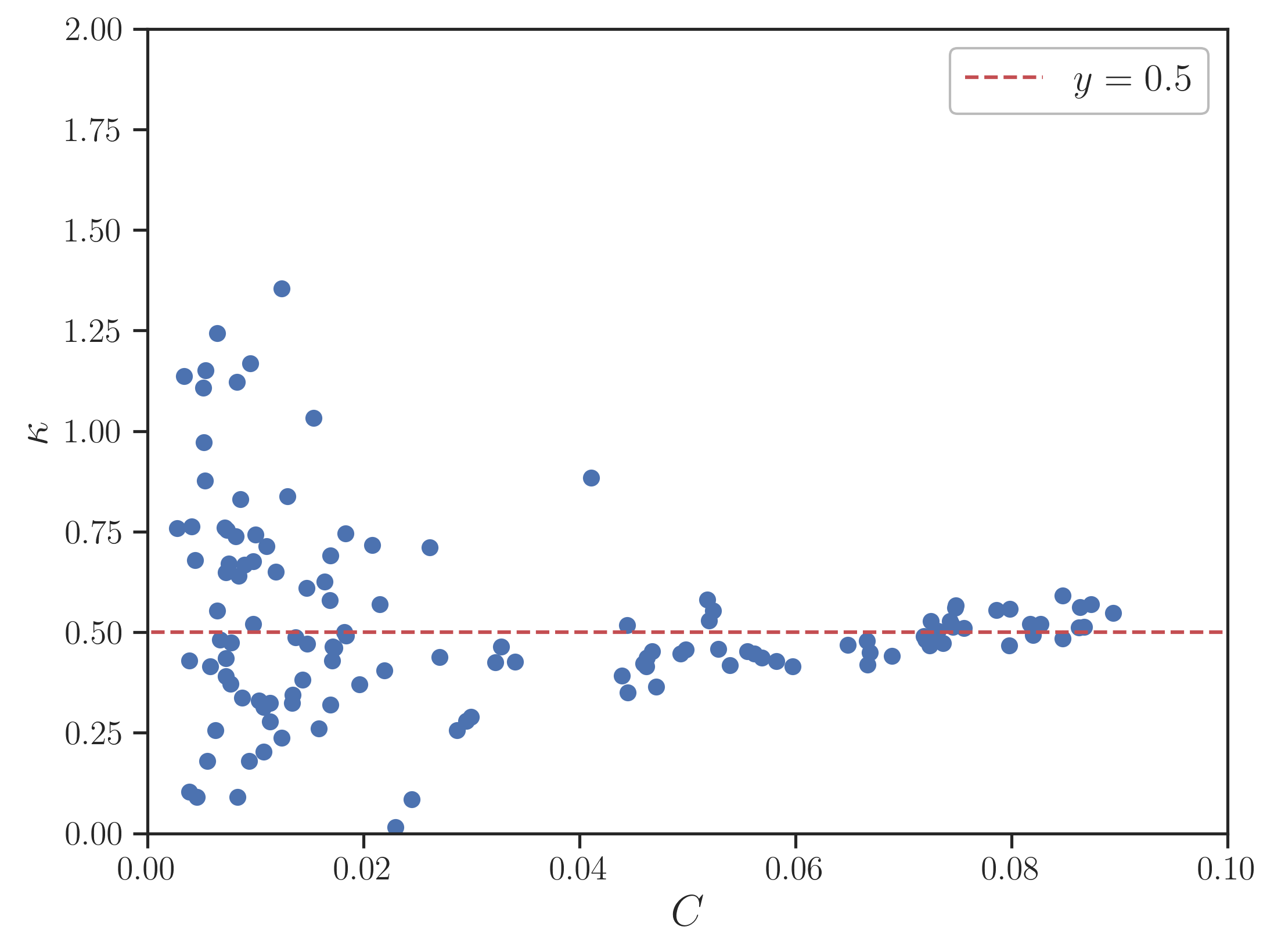}
    \caption{Coefficient of the refill function Eq. \eqref{eq:Refill_provider} when regressed separately. We used refill sequences from the top 4 most liquid stocks of our dataset, selecting the top 100 more active traders. Each dot is obtained by averaging two close-by data points, so that no individual data can be inferred from this graph.}
    \label{fig:refill}
\end{figure}

\begin{figure}[H]
    \centering
    \includegraphics[width=0.7\linewidth]{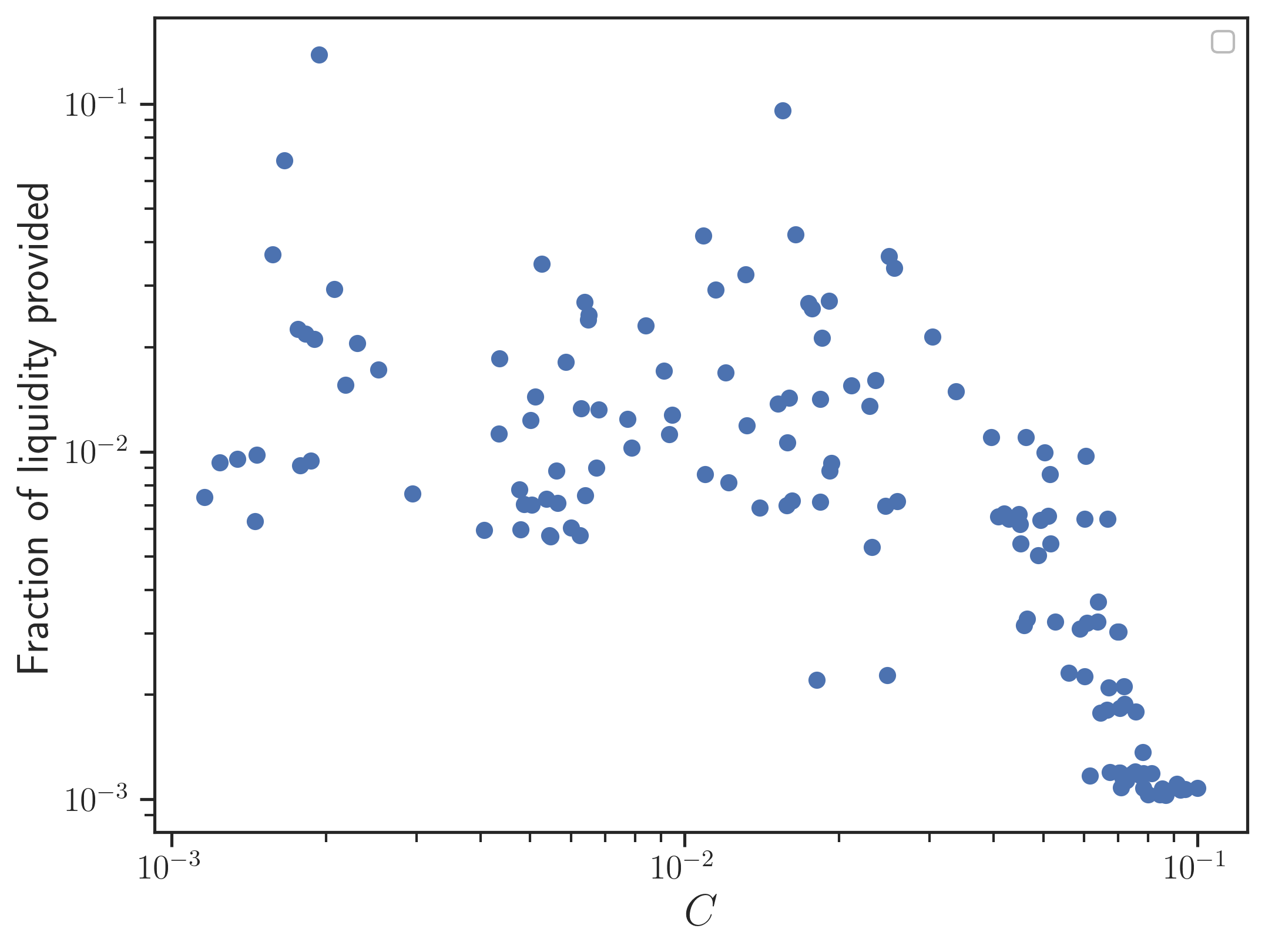}
    \caption{Plot of the fraction of liquidity provided by trader $p$ as a function of parameter $C_{\ell}$. As expected, traders responsible for most of liquidity have a low $C_{\ell}$. We used refill sequences from the top 4 most liquid stocks of our dataset, selecting the top 100 more active liquidity providers. Each dot is obtained by averaging two close-by data points, so that no individual data can be inferred from this graph.}
    \label{fig:pref_C_Coutner}
\end{figure}

We have found that the $\mathcal{K}(i)$ however depends on the liquidity provider $\ell$, some being more aggressive than others. More precisely, we fitted $\mathcal{K}_{{\ell}}(i)$ as: 
\begin{equation}\label{eq:Refill_provider}
    \mathcal{K}_{{\ell}}(i) = \frac{C_{{\ell}}}{{i}^{\kappa_{{\ell}}}},
\end{equation}
where $p$ is the label of the liquidity provider. The inverse dependence on $i$ means that, as the number of previous executions increases, liquidity providers are more willing to post competitive quotes. One can observe two main types of traders, see Fig. \ref{fig:refill}: 
\begin{itemize}
 \item High $C_{\ell} \gtrsim 0.02$ traders are ``wary'' and place their next limit order quite a bit deeper in the book. The corresponding values of $\kappa_{\ell}$ cluster around $1/2$.
\item Low $C_{\ell} \lesssim 0.02$ traders, on the other hand, correspond to ``aggressive'' liquidity providers who compete for the spread. Corresponding values of $\kappa_{\ell}$ are also larger, meaning that even after being executed many times, they are still providing competitive quotes. 
\end{itemize}
Low $C_{\ell}$ market makers are thus responsible for ensuring stable liquidity. Figure \ref{fig:pref_C_Coutner} confirms that low $C_{\ell}$ traders account for the largest fraction of consumed liquidity. Note that $C_{\ell}=0.02$ corresponds to a price degradation of $2 \%$ of the daily volatility $\sigma_D$ after the first executed limit order, and of $0.2 \%$ after the 10th execution when $\kappa_{\ell}=1$.
Interestingly, the fact that Eq. \eqref{eq:Refill_provider} decreases with $i$ suggests that the available liquidity increases away from the best price, which is the fundamental ingredient leading to a concave impact function. 

\section{Conclusion}\label{sec:conclusion}

The JPX database provides a trove of interesting features, which have only started to be exploited by Sato and Kanazawa to understand the origin of the long memory of market order signs \cite{sato2023inferring} and, more recently, to firmly establish the square-root law of market impact, Eq. \eqref{eq:sqrt_law} and rule out some of the proposed theories that predict a non-universal value for the concavity exponent $\delta$ {\cite{kanazawasato2024}}. 

Our aim in this study was to leverage the fact that {\it all} metaorders can be identified to shed light on the microscopic origin of the square-root law. Our central result, which we did not expect when starting this project, is that such a law has in fact microscopic roots, and applies already at the level of single child orders, provided one waits long enough for the market to ``digest'' these orders. This is not consistent with the standard interpretation of the LLOB model \cite{donier2015fullyconsistentminimalmodel}, which assumed that the theory described the liquidity dynamics at a ``mesoscopic'' scale. The mesoscopic impact of metaorders rather arises from a ``double'' square-root effect at the level of child orders, see Eq. \eqref{eq:sqrt_child}: square-root in volume of individual impact, followed by an inverse square root decay as a function of child order time, such that the cumulative impact of a metaorder yields back the partial (Eq. \eqref{eq:sqrt_child_partia}) and total (Eq. \eqref{eq:sqrt_Q}) square-root laws. 

This finding however immediately suggests that since market orders are anonymous, the double square-root law Eq. \eqref{eq:sqrt_child} should apply to any market orders and the impact of synthetic metaorders, reconstructed by scrambling the identity of the issuers, should also be described by the square-root impact law, Eq. \eqref{eq:sqrt_law}. We have provided empirical evidence that this both statements are indeed valid. In particular, synthetic metaorders behave exactly as real metaorders, see Fig. \ref{fig:random_japon}. We conclude that there is nothing special about child orders belonging to a given metaorder, at odds with theories that emphasize the information content of such trades to explain the square-root impact law, but in agreement with previous conjectures about the purely mechanical aspect of price impact \cite{bouchaud2010impact, toth2017short, donier2016walras}.

Interestingly, our synthetic metaorder experiment suggests that it may be possible to reconstruct the impact of metaorders from the public tape only, without trader IDs. In a forthcoming publication \cite{ustocome}, we show that this is indeed the case, provided market orders are properly aggregated into synthetic metaorders, opening the path to a new wave of empirical studies, in particular concerning cross-impact \cite{benzaquen2017dissecting, hey2023trading}. 

While our results show that the square-root impact law does not emerge at the meso-scale but is already present at the micro-scale, they trigger new unanswered questions. In particular, why is the impact of single market orders of volume $q$ also a square-root? We have argued that this is because the latent order book is locally linear, such that after a ``hot potato'' game between liquidity providers, the final counterparty of the initial market order is on average at a distance $\sqrt{q}$ from the initial mid-point. Although this scenario is intuitively plausible, we believe that a deeper dive into the JPX database (or a similar one) would allow one to (in-)validate such a picture. Another conundrum is the square-root time decay of individual market orders, which seems to violate the martingale constraint that relates the decay of the autocorrelation of the sign of market orders to impact decay \cite{bouchaud2003fluctuations, jusselin2020no}. We hope to come back to these pressing issues in the near future. 

\section*{Acknowledgments}
 We would like to thank Yuki Sato for helping with the data, and Julius Bonart, Natascha Hey, Fabrizio Lillo and Bence Toth for helpful discussions and suggestions. K. K. was supported by JSPS KAKENHI (Grant Nos.~21H01560 and 22H01141) and the JSPS Core-to-Core Program (Grant No.~JPJSCCA20200001). This research was conducted within the Econophysics \& Complex Systems Research Chair, under the aegis of the Fondation du Risque, the Fondation de l'\'Ecole Polytechnique and Capital Fund Management. We declare no financial conflict of interest. The JPX Group, Inc. provided the original data for this study without any financial support. \\
 Here we describe the author contribution to this study : GM visited Kyoto University and analyzed the TSE dataset at Kyoto University with permission from JPX, after he signed a document about the NDA strictly for academic purposes. KK supervised this research during GM's stay at Kyoto University. GL and JPB supervised this research as GM's supervisors at Ecole Polytechnique. All the authors contribute to this manuscript and agree with the findings.

\section*{Data Availability Statement}
The data supporting our results were provided by Japan Exchange (JPX) Group, Inc. JPX Group is a third-party commercial company and provided their dataset through a non-disclosure agreement with Kyoto University, strictly for academic purposes. This non-disclosure agreement imposes legal restrictions on data availability, and therefore, we cannot make the data publicly accessible without approval from JPX Group. However, upon reasonable request, KK will provide all the original data and programming code, subject to explicit approval by JPX Group.

\newpage
\printbibliography
\end{document}